\documentclass[aps,apl,reprint,a4paper,amssymb,amsmath,amsfonts,superscriptaddress,longbibliography,showkeys]{revtex4-1}
\usepackage[utf8]{inputenc}
\usepackage[T1]{fontenc}
\usepackage{lmodern}
\usepackage[american]{babel}
\usepackage{graphicx}
\usepackage{float}
\usepackage[activate={true,nocompatibility}]{microtype}
\usepackage{xcolor}
\usepackage[colorlinks=true, linkcolor=blue, citecolor=blue, urlcolor=blue]{hyperref}

\pdfoutput=1

\makeindex
\begin{document}

\title{Artificial neural networks for 3D cell shape recognition from confocal images}

\author{G. Simionato}
\thanks{These two authors contributed equally to this work}
\affiliation{Saarland University, Department of Experimental Physics, Campus E2.6 , 66123 Saarbrücken, Germany}
\affiliation{Saarland University, Department of Experimental Surgery, Campus University Hospital, Building 61.4, 66421 Homburg, Germany}

\author{K. Hinkelmann}
\thanks{These two authors contributed equally to this work}
\affiliation{Saarland University, Department of Experimental Physics, Campus E2.6 , 66123 Saarbrücken, Germany}

\author{R. Chachanidze}
\affiliation{Saarland University, Department of Experimental Physics, Campus E2.6 , 66123 Saarbrücken, Germany}
\affiliation{University Grenoble Alpes, CNRS, Grenoble INP, LRP, 38000 Grenoble, France}

\author{P. Bianchi}
\affiliation{Fondazione IRCCS Ca’ Granda Ospedale Maggiore Policlinico, Milano, Italy}

\author{E. Fermo}
\affiliation{Fondazione IRCCS Ca’ Granda Ospedale Maggiore Policlinico, Milano, Italy}

\author{R. van Wijk}
\affiliation{University Medical Center Utrecht, Department of Clinical Chemistry \& Haematology, Utrecht, The Netherlands.}

\author{M. Leonetti}
\affiliation{University Grenoble Alpes, CNRS, Grenoble INP, LRP, 38000 Grenoble, France}

\author{C. Wagner}
\affiliation{Saarland University, Department of Experimental Physics, Campus E2.6 , 66123 Saarbrücken, Germany}
\affiliation{University of Luxembourg, Physics and Materials Science Research Unit, Luxembrourg City, Luxembourg}

\author{L. Kaestner}
\affiliation{Saarland University, Department of Experimental Physics, Campus E2.6 , 66123 Saarbrücken, Germany}
\affiliation{Saarland University, Theoretical Medicine and Biosciences, Campus University Hospital, Building 61.4, 66421 Homburg, Germany}

\author{S. Quint}
\email[]{Stephan.Quint@physik.uni-saarland.de}
\affiliation{Saarland University, Department of Experimental Physics, Campus E2.6 , 66123 Saarbrücken, Germany}
\affiliation{Cysmic GmbH, Geretsrieder Str. 10, 81379 M\"unchen, Germany}

\date{\today}

\begin{abstract}
We present a dual-stage neural network architecture for analyzing fine shape details from microscopy recordings in 3D. The system, tested on red blood cells, uses training data from both healthy donors and patients with a congenital blood disease. Characteristic shape features are revealed from the spherical harmonics spectrum of each cell and are automatically processed to create a reproducible and unbiased shape recognition and classification for diagnostic and theragnostic use.
\end{abstract}

\maketitle
\newpage
\section{Main}
Cell morphology is a phenotypic characteristic reflecting the cell cycle, metabolic state or cellular activity \cite{wu2020single,cutiongco2020predicting,prasad2019cell}. While brightfield imaging is affected by the orientation of cells on the microscopy slide, which determines a certain projection, 3D confocal microscopy allows to investigate the whole cell surface without loss of information. 

The analysis of shapes is related to feature detection in processed images. Deep-learning-based approaches can potentially be employed for such tasks, avoiding manual procedures that are time consuming, subjective and prone to human error. For this reason, we developed a method that provides recognition at high shape-detail resolution of 3D objects that are similar in shape and nature, thus having the potential to be universally used due to its precision. This method is implemented with a low computational cost. 

To evaluate our system, we employed red blood cells (RBCs), representing one the most deformable cell types. In healthy subjects, RBCs in stasis are typically biconcave disks, but external factors such as the pH or osmolarity of the suspension medium or interaction with surfaces can stimulate a shape transition. Such transformations appear in a distinct order and are described as the stomatocyte-discocyte-echinocyte (SDE) sequence \cite{lim2008}. In case of blood diseases, additional morphological abnormalities appear, defining certain blood disorders (e.g., hereditary spherocytosis, sickle cell disease, acanthocytosis or stomatocytosis) \cite{diezsilva2010, hardie1989acanthocytosis}. The investigation of RBC morphology for the diagnosis of hematological diseases relies on the visual examination of blood smears. Advances in automation of the analysis have especially involved convolutional neural networks (CNNs) for white blood cell recognition \cite{sahlol2020efficient} and, in some cases, for RBC detection and shape classification in 2D, both in stasis and flow \cite{yao2019cell,kihm2018classification,xu2017deep}. However, in blood smear preparation the smearing and drying procedures affect cells, leading to unwanted morphological deformation \cite{wenk1976comparison} and loss of the 3D information of the original cell shape. 

\begin{figure*}[t!]
	\centering
	\includegraphics[width=18cm]{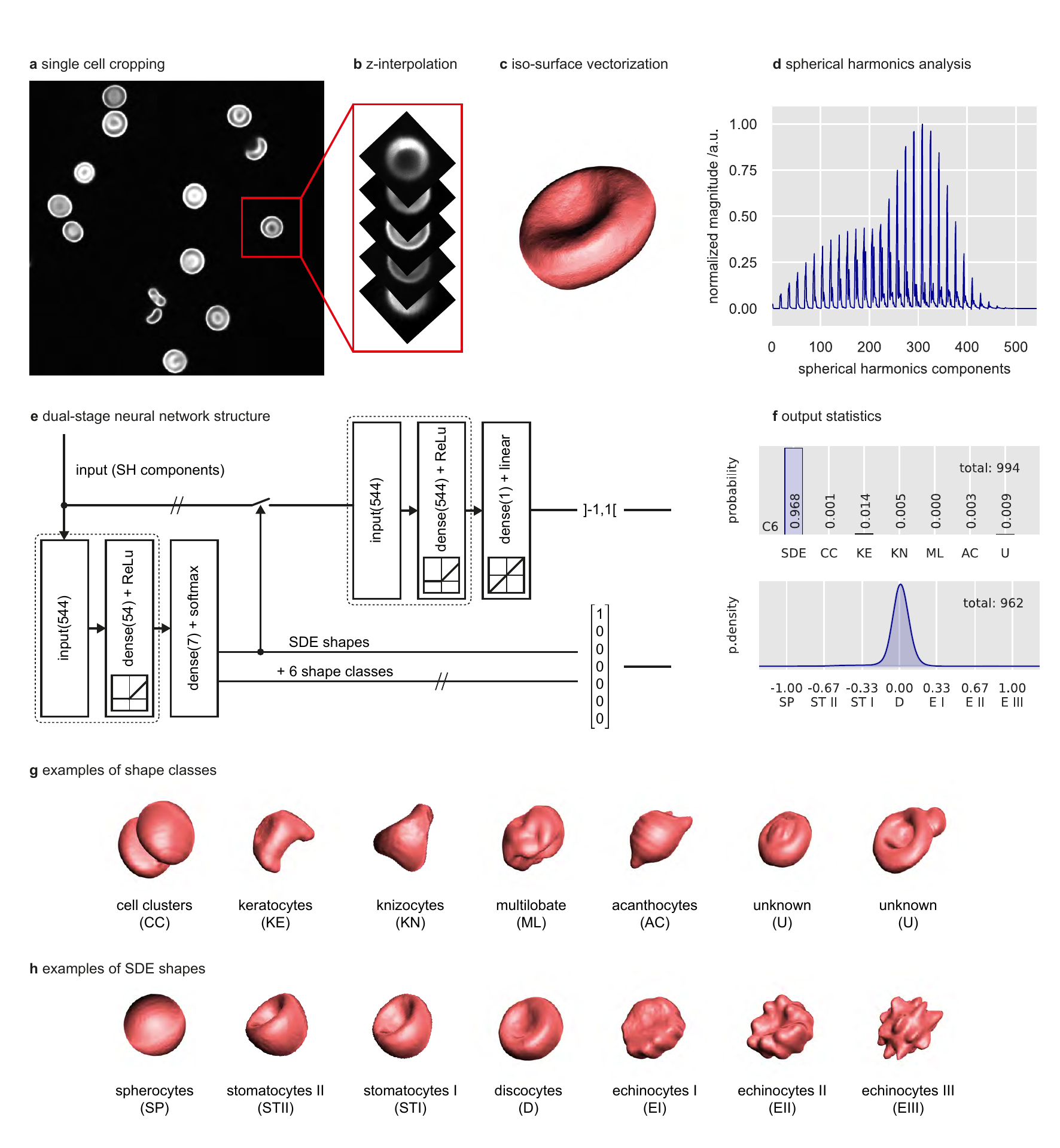}
	\caption{\textbf{Workflow for automatic classification by the dual-stage ANN.} \textbf{a} After sample staining and imaging by confocal microscopy, each cell is cropped individually and the full stack is interpolated in the z direction to achieve isotropic resolution (\textbf{b}). \textbf{c} The isosurface is retrieved by applying a constant threshold to each cell. \textbf{d} The vectorized data are transformed with respect to their spherical harmonics ($L_{2}$-norm of $32$ radii and corresponding $16$ frequencies, see Methods), representing a rotation invariant form of the cell shape. \textbf{e} Data are fed to the dual-stage ANN, with the first-stage resulting in a classification output (bottom). This stage consists of a three-layer architecture providing an input layer ($544$ neurons), a fully connected hidden layer with a ReLU activation function ($54$ neurons) and a fully connected output layer (softmax activation). The output is represented by a vector of size $7$, which is subjected to discrimination by a given threshold. If the confidence is higher than $75\,\%$ (threshold), the output is assigned to a certain existing class. Otherwise, the output is assigned to unknown (U) shapes. All detected SDE shapes are forwarded to the second-stage ANN, with an anatomy similar to that of the first ANN, except for the hidden layer that exhibits $544$ neurons. The regression-type output layer assigns each cell a score between $-1$ and $+1$ and allocates them on the continuous SDE scale (top). \textbf{f} Example of a typical resulting shape distribution in a healthy subject. Almost all RBCs are discocytes. \textbf{g} Representative images for $6$ mutually exclusive shape classes (see Methods for description): two out of many different examples of unknown shapes are shown. \textbf{h} Representative cells of the SDE scale. The training data included SDE shapes artificially induced by changing the osmolarity of the suspension medium.}
	\label{fig1}
\end{figure*}

Instead of this technique, we performed fixation of RBCs, followed by fluorescent staining. Confocal microscopy was used to capture the 3D representations of cells by means of z-stacks (Fig. \ref{fig1}a). After offset elimination, intensity normalization and adaptation of the resolution in the x/y and z direction by interpolation (Fig. \ref{fig1}b), we discriminated the cell membrane as an isosurface defined by a constant intensity threshold (Fig. \ref{fig1}c). In contrast to preexisting classification approaches for such kinds of data, e.g., 3D-CNNs \cite{kamnitsas2017efficient} or voxelwise processing techniques \cite{qi2016volumetric}, we transformed and subsequently collapsed the volumetric data to exclusively access the features of interest. This data reduction was achieved by decomposing the spatial information of the cell surface into the respective spherical harmonics (SH) spectrum (Fig. \ref{fig1}d) \cite{kazhdan2003rotation}. Thus, a one-dimensional data-vector was obtained, encoding the prevalent features of the 3D shape and characterized by rotation and translation invariance (see Methods). A subsequent normalization mapped the SH spectrum into the range from 0 to 1, rendering the data suitable to train artificial neural networks (ANNs).

For cell shape recognition, we used a dual-stage ANN architecture (Fig. \ref{fig1}e). The first stage was designed to sort out distinct RBC shapes that do not fit the SDE spectrum. Such shapes particularly occur in samples from patients with blood diseases or other pathologies \cite{diezsilva2010, hardie1989acanthocytosis}. An additional class of “unknown” cells was added to reflect human uncertainty regarding unclear or rare shapes not yet defined in the literature. This class included all cells classified by the first-stage ANN with an identification accuracy below a threshold of 75\%.

The second stage served to discriminate all SDE shapes. Previously, the SDE sequence was described by assigning different shapes to pseudodiscrete classes, such as spherocytes, stomatocytes type I/II, discocytes and echinocytes type I/II/III, to serve as reasonable support for manual classification \cite{lim2008}. By employing supervised training for both ANNs, we used the state-of-the art classification scheme to create training data for carefully selected sets of non-SDE (Fig. \ref{fig1}g) and ideal SDE (Fig. \ref{fig1}h) shapes.

In between the pseudodiscrete SDE classes, extra transition shapes were observed. For this reason, we assumed that the shape transformation occurs in a continuous manner and introduced a linear scale to automatically assign any identified SDE shape to an interval ranging from $-1$ (spherocytes) over $0$ (discocytes) to $+1$ (echinocytes). The regression-type ANN of the second stage allowed for fine distinction of morphological details within the whole spectrum at a precision that is manually unattainable.

The overall system consisted of (1) a classification-type ANN, which assigns each cell to one out of seven types (SDE shapes, knizocytes, keratocytes, acanthocytes, multilobate cells, cell clusters and unknown cells, Fig. \ref{fig1}f top), followed by (2) a regression-type ANN, which characterizes all detected SDE shapes (spherocytes, stomatocytes type II/I, discocytes and echinocytes type I/II/III, Fig. \ref{fig1}f bottom).

The need for a vast amount of data for both training and validation was met by exhaustive augmentation. This was performed by selecting random pairs of cell shapes followed by random superposition and normalization of their spectra. In particular, we created artificial data within each of the mutually exclusive classes for the first-stage ANN as well as random interpolations between neighboring or the same SDE pseudodiscrete classes for the second-stage ANN. Avoiding a dependency between training and validation data, the split was performed before interpolation (for training and validation loss and accuracy see Supplementary Fig. 1 and 2).

\begin{figure*}[t!]
	\centering
	\includegraphics[width=14cm]{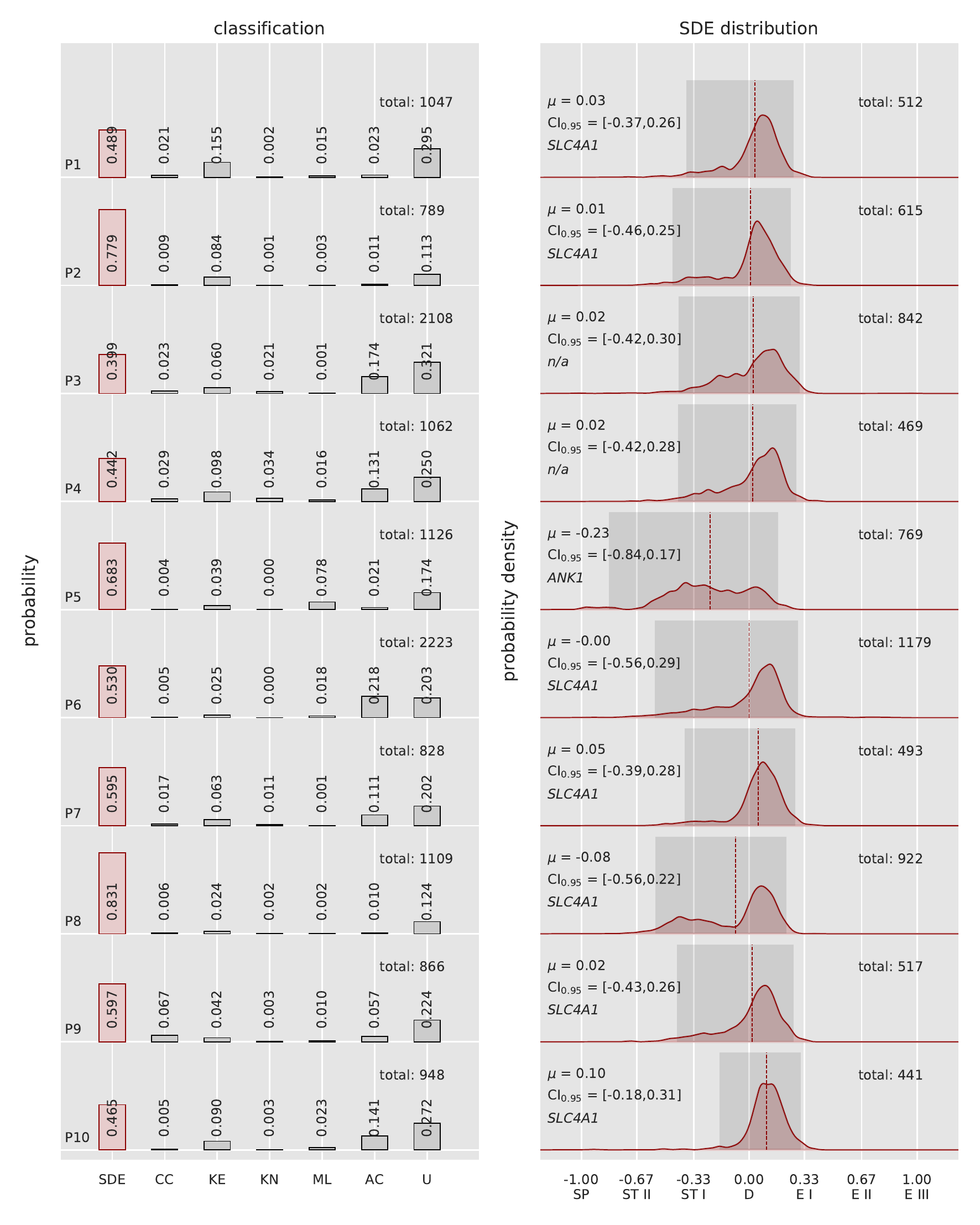}
	\caption{\textbf{Automatic 3D RBC shape recognition for patients affected by hereditary spherocytosis.} The total number of cells  analyzed in both the classification and SDE distribution per patient are indicated. The probability density distributions within the SDE range show the expected value $\mu$ (dashed red line) and highlight 95\% confidence intervals (dark gray area). The results demonstrate that most of the patients have expected values related to discocytes, with a tendency toward stomatocytes. A population of spherocytes (score $-1$) is lacking in all $10$ samples, proving that such a shape is not a hallmark of the disease. Additionally, shape profiles among patients are different, suggesting a relation to the various genetic mutations (\textit{SLC4A1}, and in one case \textit{ANK1}). In particular, P1 and P2 as well as P3 and P4 (nonidentified mutation) are relatives and show similar profiles, especially within the SDE range. P5 harbors a mutation that affects the cytoskeletal protein ankyrin, resulting in the highest number of stomatocytes, including some spherocytes. P6, P7 and P9 are affected by mutations in the band 3 protein, as is P8, who also has a double mutation in spectrin alpha (see Supplementary Table 1), although this latter mutation is not known to be pathogenic. The differences among these patients may depend on the particular mutation altering the same gene: P6 showed 22\% acanthocytes, which occur in variable numbers in P1 and P2, P7 and P9. The phenotypic associated defect in fact, in some cases, causes band 3 deficiency, while in others, it leads to spectrin deficiency (Supplementary Table 1). Other classes showed rare occurrences, and shape deviations were classified as unknown cells in all the tested samples, suggesting that a larger amount of shape deformations were detected by the ANNs.}
	\label{fig2}
\end{figure*}

Our system was validated through the inspection of blood samples from 10 healthy donors (Supplementary Fig. 3), where a prevalence in discocytes is expected. Then, the method was tested on 10 patients with the most commonly mutated genes causing hereditary spherocytosis (\textit{SLC4A1}, \textit{ANK1}, Fig. \ref{fig2}, Supplementary Note and Supplementary Table 1). As highlighted in a confusion matrix created for healthy controls and patients (Supplementary Fig. 4), the comparison of manual classification and automatic classification (first-stage ANN) resulted in a variable mismatch due to the rare occurrence of related shapes in blood samples and limited available training data. On the other hand, we observed a high agreement between the automatic allocation of cells on the SDE scale (second stage), ranging from 78\% to 100\%.

From a clinical perspective, patients' data resulted in a different statistical output compared to those of healthy donors (Fig. \ref{fig1}f and Supplementary Fig. 3), with wider shape distributions within the SDE scale and an expected value toward stomatocytes for P5 (Fig. \ref{fig2}). The tendency of RBCs to form a round shape is a hallmark of hereditary spherocytosis, and spherocytes are particularly expected in blood smears of affected subjects. A previous study reported that 2.6\% of blood smears in a set of 300 patients did not exhibit detectable spherocytes, leading to a possible misdiagnosis \cite{mariani2008clinical}. However, the evaluation in 3D indicated that spherocytes in the tested set of patients were very rare or completely absent and comparable in number to those found in healthy subjects. These results confirmed the dependence of blood smear shape analysis on cell rotation (Fig. \ref{fig2}, Supplementary Fig. 5), proving that spherocytes are not the main shape in hereditary spherocytosis.  

Some indications of the presence of other shape deformations in blood smears were reported \cite{perrotta2008hereditary, eber2004hereditary} and may have an association with the different genetic mutations causing the disease. The fine recognition of shape details by the automated dual-stage ANN resulted in a differential shape profile for various mutations. This represents additional information compared to that obtained from blood smears, where solely the type of blood disease can be discriminated. Finally, the prevalence of shapes occurring in non-SDE classes, especially in the unknown class, underlined the high morphological variability in patients, highlighting the demand for further RBC shape definitions.
 
In conclusion, the proposed approach describes an automated evaluation system for cell morphology in addition to or instead of manual methods. Its application in hematology revealed that conventional microscopy has limitations with regard to cell morphology that may lead to erroneous interpretations and shows the superiority of 3D visualization and characterization of cell shapes. In addition to the unbiased outcome, automation by ANN allowed both the recognition of small shape details and the possibility of using a regression-based approach for cells undergoing continuous shape transitions. Owing to the details revealed using 3D imaging combined with ANNs as a universal tool for shape recognition, thorough tests on anemic subjects may render our method suitable for diagnostic purposes. In addition, from the results obtained with the tested pool of patients, we observed potential applicability with larger datasets to relate the ANN output to a particular mutation. While genetic analysis is the gold standard for detection, cell imaging can be of additional interest in the investigation of the severity and state of a disease \cite{kaestner2020trends}. Moreover, it can be applied for personalized theragnostics when the effectiveness of a specific treatment is tracked \cite{alvarez2015quantification}. Our method may be adapted for other cell biological applications or even industrial purposes.

%\clearpage
%\bibliography{references2}
\clearpage
\section{ACKNOWLEDGMENTS}
This work was supported by the Volkswagen Experiment! grant, the Deutsche Forschungsgemeinschaft (DFG) in the framework of the research unit FOR 2688 and the European Union’s Horizon 2020 Research and Innovation Programme under the Marie Skłodowska-Curie grant agreement no 860436 – EVIDENCE.

\section{AUTHOR CONTRIBUTIONS}
G.S. performed the experimental procedures, microscopy, training data generation by manual classification and manuscript writing. K.H. performed 3D rendering of cells, video editing, programming, process automation and system validation. R.C. performed preprocessing of microscopy images, additional data acquisition and reviewing of the manuscript. P.B., E.F. and R.W. performed diagnostic analysis, data collection and patient blood sampling. M.L. and C.W. discussed the results; C.W. provided laboratory infrastructure and consumables for experiments. L.K. and S.Q. designed the project. S.Q. conceived the method; performed programming and system optimization; supervised data evaluation, validation, manuscript writing; and acquired funds. All authors contributed to the editing and proofreading of the manuscript.

\section{COMPETING INTERESTS}
The authors declare no competing interests.

\section{ADDITIONAL INFORMATION}
Supplementary information available.

\section{DATA AVAILABILITY}
The raw data supporting findings of this study are available at: \url{https://gir1.de/cytoShapeNet/data.html}.

\section{CODE AVAILABILITY}
The code is open source under GNU General Public License (GPL), available on GitHub at: \url{https://github.com/kgh-85/cytoShapeNet}.

%\bibliography{Simionato_et_al_2020}

%%merlin.mbs apsrev4-1.bst 2010-07-25 4.21a (PWD, AO, DPC) hacked
%%Control: key (0)
%%Control: author (8) initials jnrlst
%%Control: editor formatted (1) identically to author
%%Control: production of article title (-1) disabled
%%Control: page (0) single
%%Control: year (1) truncated
%%Control: production of eprint (0) enabled
%

\end{document}

\large
\begin{center}
\textbf{Supplementary information}
\end{center}

\title{Supplementary}
\maketitle

\clearpage
\begin{figure}[h!]
	\centering
	\includegraphics[width=17cm]{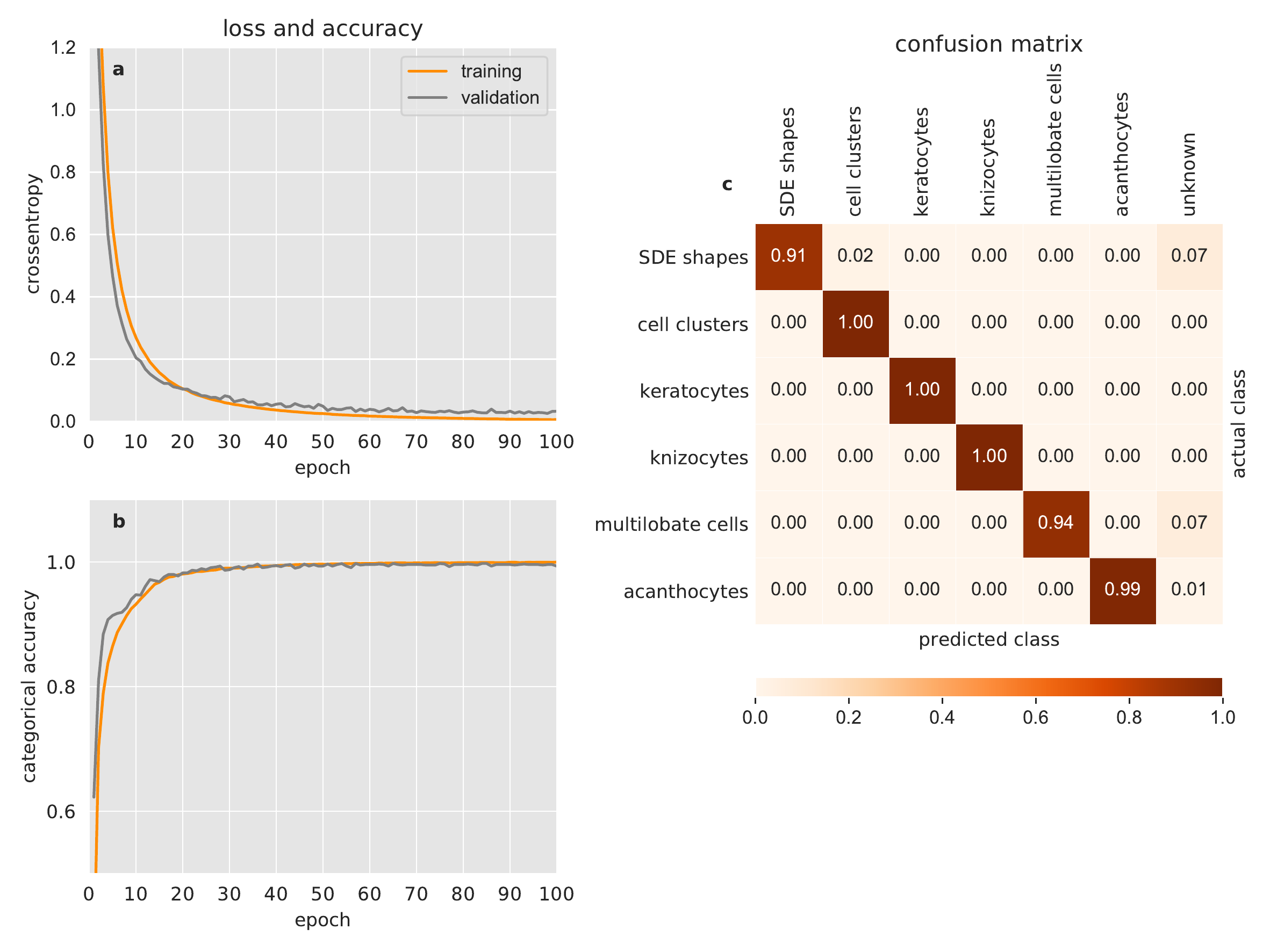}
	\caption{\textbf{Training and validation loss and accuracy for the first-stage ANN.}
		\textbf{a} During training, the employed loss function (crossentropy) was minimized throughout $100$ epochs. \textbf{b} The categorical accuracy of cell classification for the training and validation sets converged close to $100\,\%$. The validation split was $20\,\%$. \textbf{c} The accuracy was further evaluated by means of a confusion matrix. “Predicted” versus “actual” (manually classified) cells demonstrate very good concordance, ranging from $91\,\%$ to $100\,\%$.}
	\label{fig1}
\end{figure}

\clearpage
\begin{figure*}[t!]
	\centering
	\includegraphics[width=17cm]{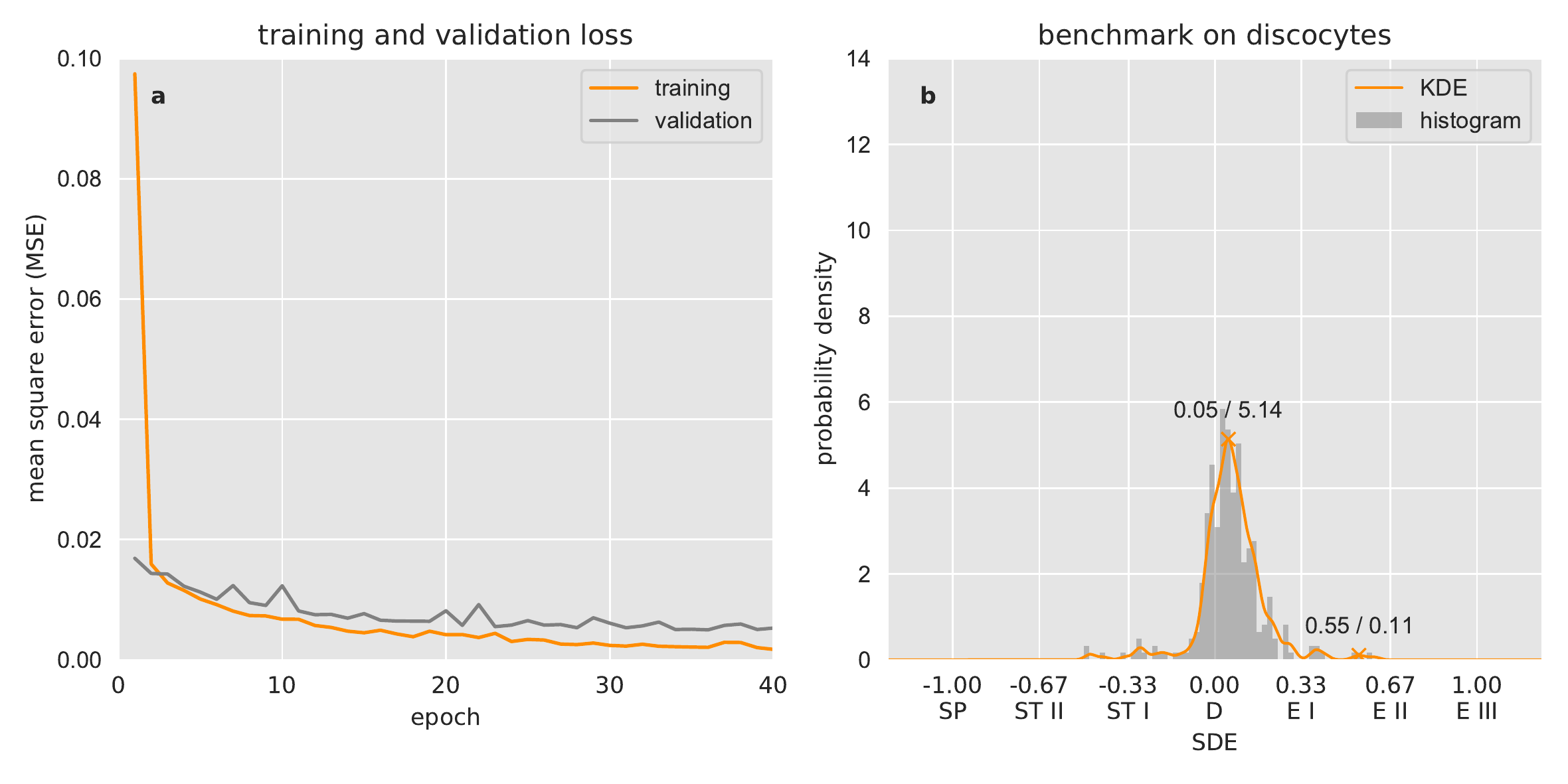}
	\caption{\textbf{Training and validation loss for the second-stage ANN.} \textbf{a} Within $40$ epochs, the mean squared error (MSE) of the system was minimized. \textbf{b} Test on a representative set of $308$ independent discocytes. The head of the distribution is allocated at $0.05$, and the vast majority of cells is located within the range of $-0.15$ to $+0.15$.}
	\label{fig2}
\end{figure*}

\clearpage
\begin{figure*}[t!]
	\centering
	\includegraphics[width=14cm]{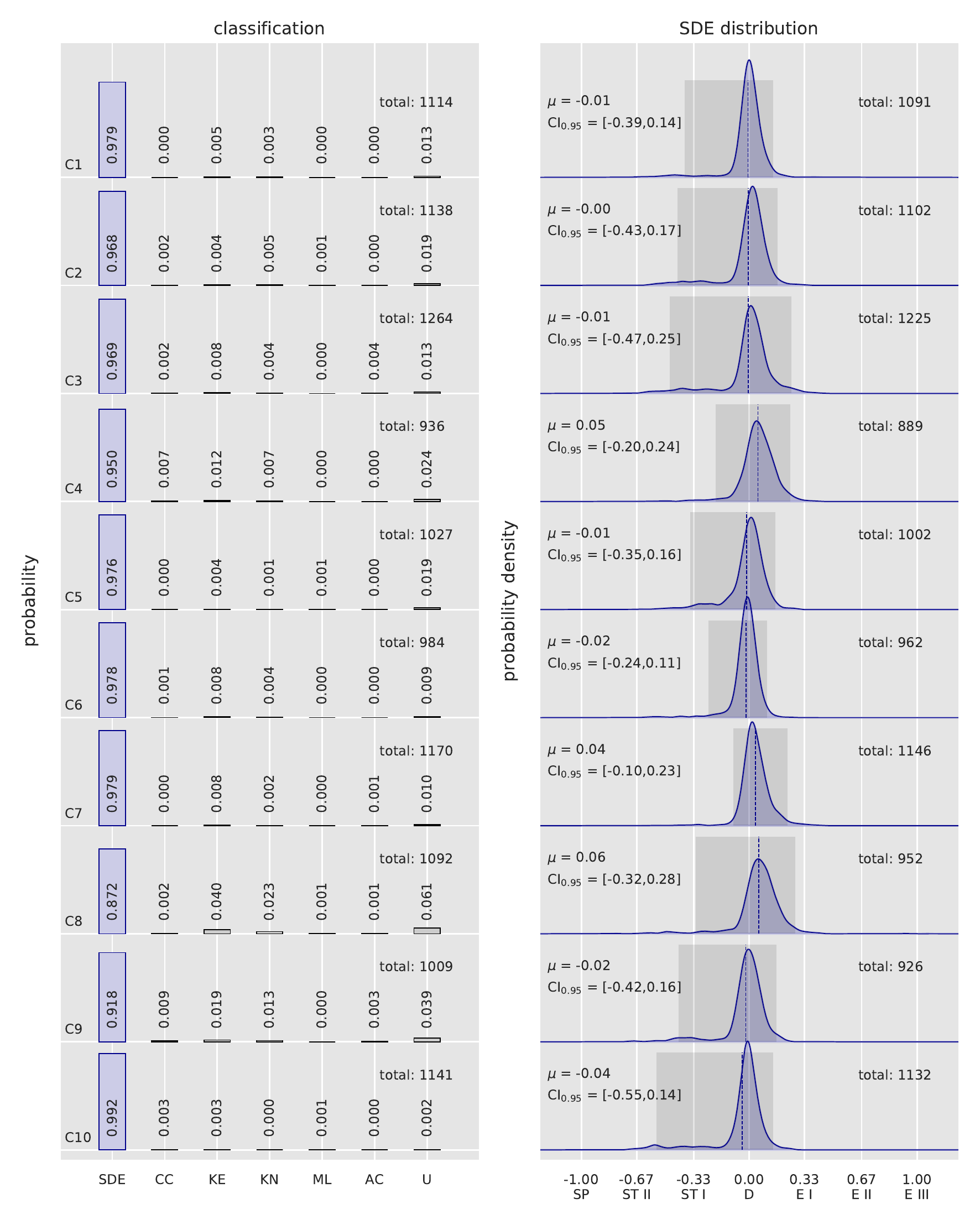}
	\caption{\textbf{Automatic 3D shape recognition of RBCs from 10 healthy subjects.} As expected, in healthy individuals, shape distributions are centered around discocytes ($0.00$), with expected values $\mu$ ranging from $-0.04$ to $+0.06$. The $95\,\%$ confidence interval ($\text{CI}_{0.95}$) demonstrates that shape distributions do not involve echinocytes, except for C8, which extends to the range of echinocytes I ($+0.33$). A very small amount of stomatocyte I exists in samples from most donors. The vast majority of cells are identified as “SDE shapes” (see total number in the classification panel versus total in the SDE distribution). The classification ANN contains a small percentage of cells classified as unknown. These results support the reproducible outcome of the dual-stage ANN analysis of RBCs from different donors and corroborate the differences seen between patients.}
	\label{fig3}
\end{figure*}

\clearpage
\begin{figure*}[t!]
	\centering
	\includegraphics[width=14cm]{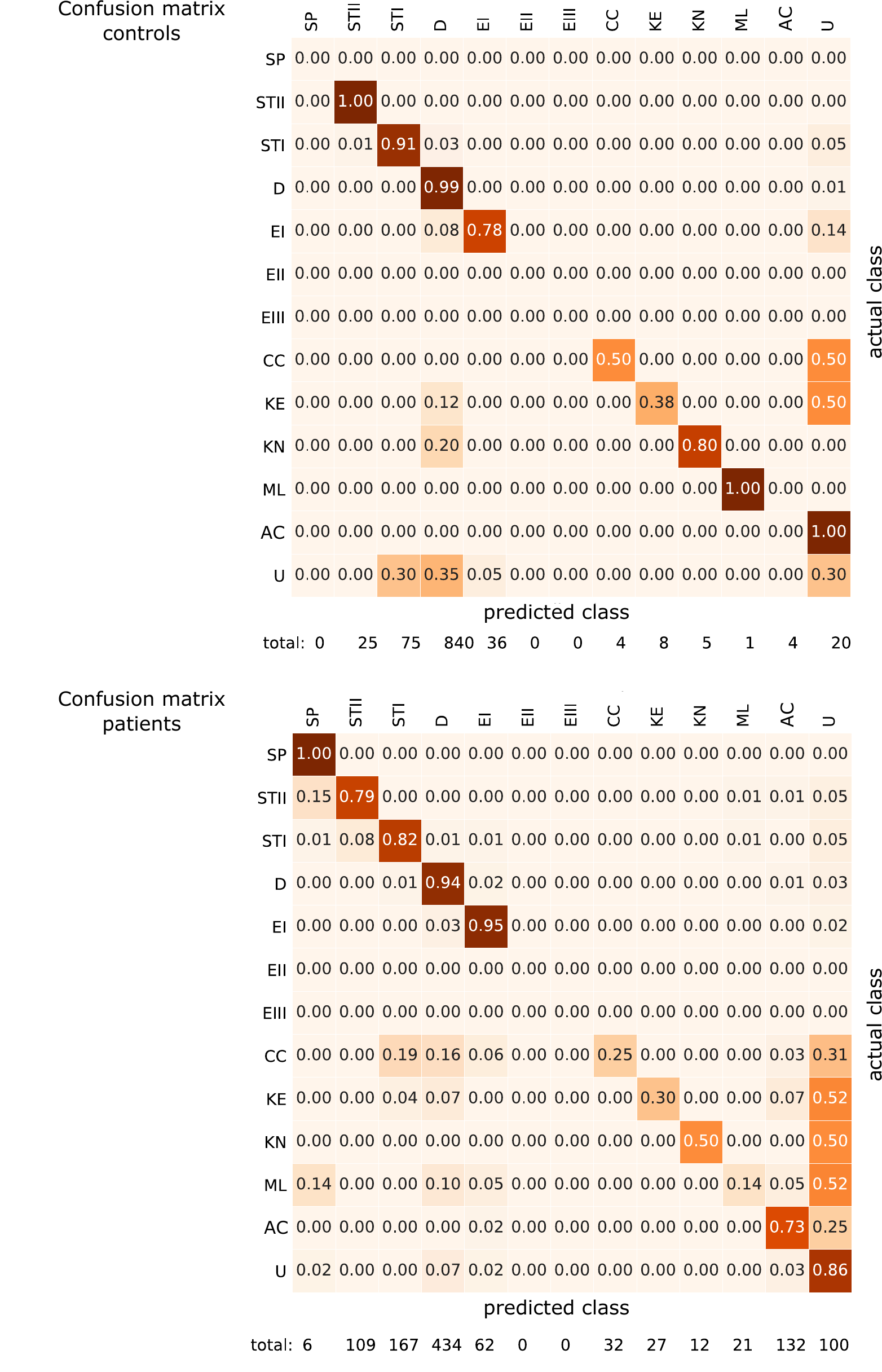}
	\caption{\textbf{Confusion matrices for healthy controls and patients with hereditary spherocytosis comparing “predicted” and “actual” (manually classified) shape classes.} SDE shapes resulted in excellent recognition accuracy, ranging from $79\,\%$ up to $100\,\%$, based on $2000$ randomly picked cells from $5$ healthy subjects and $5$ patients ($10\,\text{donors}\times 200$ cells) that were not included in the training process. Cells belonging to specific classes showed a higher recognition error, partly due to the higher amount of unknown shapes, especially those occurring in patients, partly due to the less well-defined shape features that allow us to clearly distinguish each class. The observed error in such classes is related to the intrinsic inaccuracy of the training data.}
	\label{fig5}
\end{figure*}

\clearpage
\newpage
\section*{Note on hereditary spherocytosis and tested patients}
Hereditary spherocytosis is the most common hemolytic anemia in subjects of northern European ethnicity. This disease is mostly inherited in an autosomal dominant manner, although in approximately $20\,\%$ of cases, inheritance is autosomal recessive or due to \textit{de novo} mutations \cite{bridges2008anemias}. The mutated genes code for RBC cytoskeletal proteins, most commonly the anion exchanger band 3 and the cytoskeletal protein ankyrin, as well as spectrin and proteins 4.1 and 4.2. Occasionally, patients are affected by a combination of multiple mutations in these genes. The defect mostly translates into a deficiency of the mutated protein. However, the genetic mutation may result in the deficiency of a different protein. An example is defects in band 3 and ankyrin occasionally causing the lack of spectrin integration in the membrane and degradation of free spectrin molecules, resulting in spectrin deficiency (\cite{perrotta2008hereditary, iolascon1283}, Supplementary Table \ref{table1}). Different mutations may occur in the same gene, leading to variations in disease severity \cite{dhermy1997heterogenous}.\\

Mutations in the abovementioned genes eventually lead to RBC membrane vesiculation that reduces the cell surface-to-volume ratio, transforming the RBC from a discocyte to a sphere-like shape \cite{bernhardt2013red}. These RBCs are named spherocytes and typically appear in blood smears as cells with a smaller and circular projected area, devoid of the characteristic central pallor observed in discocytes (\ref{fig5}). Such RBCs are less deformable, resulting in a reduced lifespan, which may eventually lead to anemia. This latter may be severe, moderate, mild or even absent when RBC loss is balanced by enhanced erythropoiesis.\\

Typical complications involve splenomegaly, reticulocytosis and hemolytic anemia, which can require exchange transfusions \cite{hassoun1996hereditary}. The only existing treatment is splenectomy, which improves cell survival and, reduces anemia, reticulocyte count and hyperbilirubinemia, but not the presence of spherocytes. Additional prophylaxis against infections is recommended \cite{eber2004hereditary}.\\

The variable symptomatology and the numerous mutations make hereditary spherocytosis a highly heterogeneous disease. Moreover, spherocytes can be observed in other diseases \cite{deng2015misdiagnosis} and can also be present as an artifact of the blood smear technique. Therefore, establishment of the correct diagnosis is dependent on several different tests.\\

For the $10$ patients presented in this work, the diagnostic criteria were based on the following evaluations: presence of chronic hemolytic anemia, RBC morphology examination on blood smears, eosin-5'-maleimide (EMA) binding test, osmotic fragility test or altered osmotic gradient ektacytometry curve. Confirmation tests included sodium dodecyl sulfate-polyacrylamide gel electrophoresis (SDS-PAGE) analysis of RBC membrane proteins and  next-generation sequencing (NGS) for mutation identification. The mutations and related detected protein defects are summarized in Table \ref{table1}. Patients P1 and P2 are relatives, as are P3 and P4. The latter were diagnosed with hereditary spherocytosis, but their mutation could not be detected. All patients were heterozygous for their main mutation. P8 was additionally found to be heterozygous for a point missense mutation (Supplementary Table \ref{table1}) and homozygous for the $\alpha$ LELY low-expression alpha-spectrin variant. The missense mutation is considered a variant of uncertain significance (VUS), similar to homozygosity for LELY, but the latter may contribute to the eventual defective protein expression. The band 3 mutation is likely pathogenic and the main cause of the disease.

\clearpage
\begin{figure}[t!]
	\centering
	\includegraphics[width=14cm]{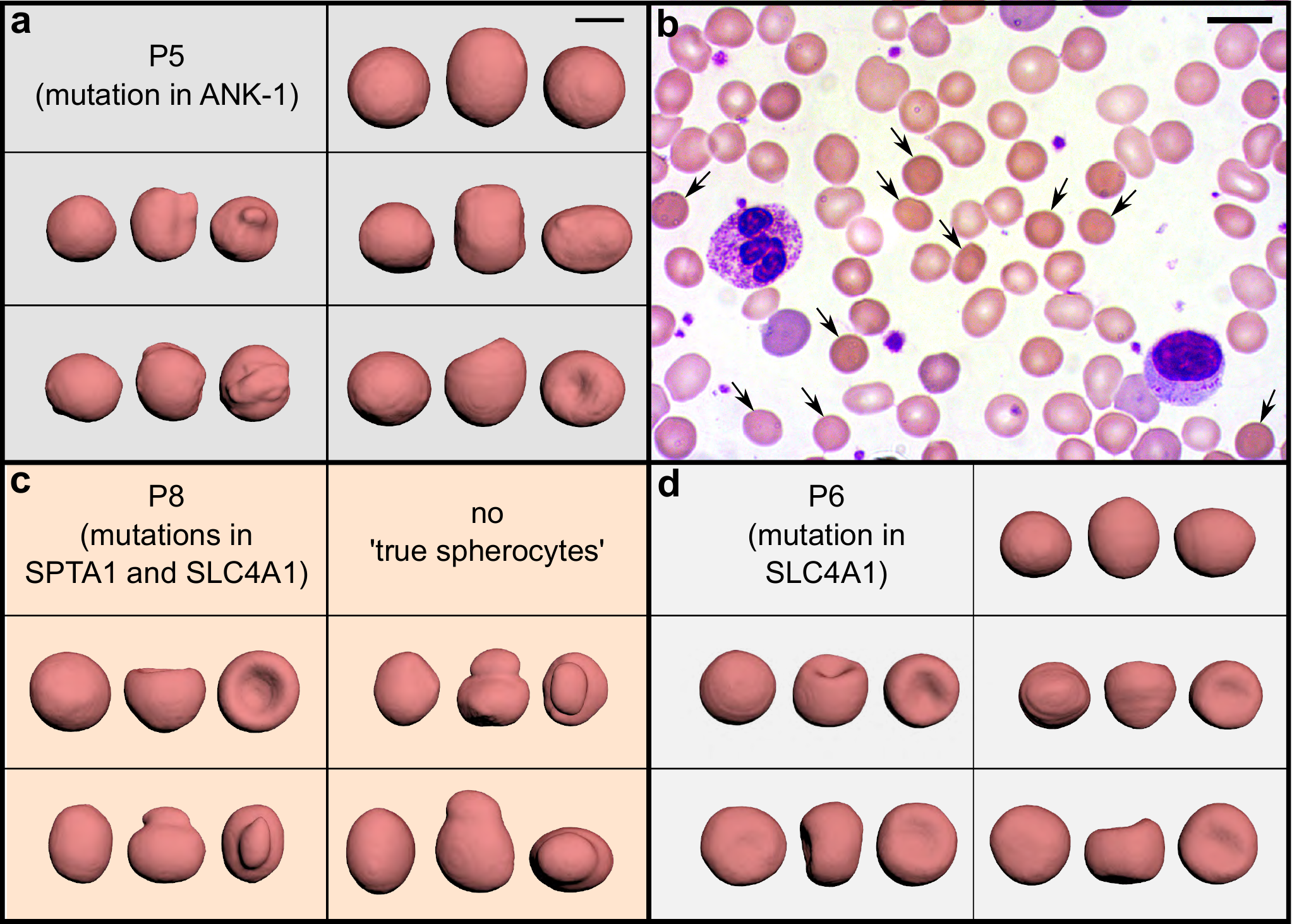}
	\caption{\textbf{Typical RBCs from $3$ patients affected by hereditary spherocytosis.} Each box shows three different rotations of the same cell; scale bar=$4\,\mu\text{m}$. \textbf{a} RBCs from a patient with a mutation affecting ankyrin-1 and the respective blood smear in \textbf{b}; scale bar=$10\,\mu\text{m}$. While several spherocytes appear on the smear (arrows), 3D reconstructions show different kinds of shapes. Top boxes: mutated proteins are indicated. The top right box in each panel shows a “true” spherocyte from 3 different viewing angles. \textbf{c} and \textbf{d} are patients affected by other mutations. No “true” spherocytes were observed in 3D in \textbf{c}. \textbf{d} Patient with a mutation in band 3, mostly showing stomatocytes rather than spherocytes.}
	\label{fig6}
\end{figure}

\clearpage
\newpage
\begin{table}[h!]

	\begin{center}
		\caption{Information on the tested hereditary spherocytosis patients.}
		\label{table1}
		\begin{tabular}{c|c|c|c|c}
			\multirow{2}{*}{\textbf{Patient}} & \multirow{2}{*}{\textbf{Mutated gene}}& \multicolumn{1}{>{\centering\arraybackslash}m{40mm}|}{\textbf{Corresponding}}& \multirow{2}{*}{\textbf{Mutation}}& \multicolumn{1}{>{\centering\arraybackslash}m{50mm}}{\textbf{Phenotypical}}\\
			& & \multicolumn{1}{>{\centering\arraybackslash}m{40mm}|}{\textbf{protein}} & & \multicolumn{1}{>{\centering\arraybackslash}m{50mm}}{\textbf{defect}} \\
			\midrule
			\midrule
			P1 & \textit{SLC4A1} & band 3 & c.2423G>A (p.R808H) & spectrin deficiency \\
			& & & & \\
			P2 & \textit{SLC4A1} & band 3 & c.2423G>A (p.R808H) & spectrin deficiency \\
			& & & & \\
			P3 & not identified & n.a. & n.a. & spectrin deficiency\\
			& & & & \\
			P4 & not identified & n.a. & n.a. & spectrin deficiency \\
			& & & & \\
			P5 & \textit{ANK-1} & ankyrin-1 & c.2559-2A>G (splicing) & predicted skipping of exon 26\\ 
			& & & & and frameshift, loss \\
			& & & & or truncated ankyrin \\
			& & & & \\
			P6 & \textit{SLC4A1} & band 3 & c.620delG (p.G207fs) & loss or truncated band 3 \\
			& & & & \\
			P7 & \textit{SLC4A1} & band 3 & c.2279G>A (p.R760Q) & band 3 deficiency \\
			& & & & \\
			P8 & \textit{SLC4A1}, & band 3, & c.2279G>A (p.R760Q) & band 3 deficiency \\
			& \textit{SPTA1} & spectrin & c.3841C>T (p.R1281C) & \\
			& & and $\alpha$ LELY & & \\
			& & & & \\
			P9 & \textit{SLC4A1} & band 3 & c.163delC (p.H55TfsX11) & band 3 deficiency \\
			& & & & \\
			P10 & \textit{SLC4A1} & band 3 & c.2510C>A (p.T837K) & band 3 deficiency \\
			
		\end{tabular}
	\end{center}
\end{table}

\clearpage
\begin{figure*}[h!]
	\centering
	\includegraphics[width=14cm]{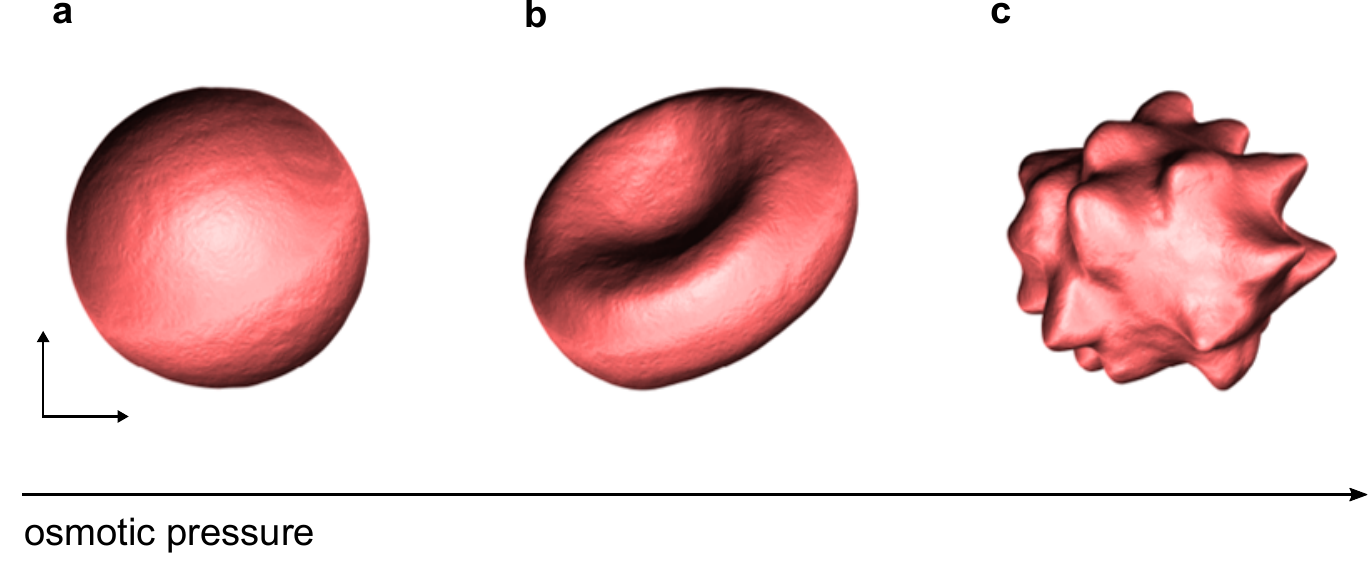}
	\caption{\textbf{SDE training data were obtained by exposing RBCs from healthy donors to a different osmotic pressure.} In isotonic solution, most of the RBCs are discocytes, \textbf{b}. Upon decreasing the osmotic pressure (hypotonic solution), the RBCs exhibit swelling and transform into stomatocytes and further into spherocytes, \textbf{a}. On the other hand, echinocytes develop in hypertonic solution, \textbf{c}. Scale bar=$2\,\mu\text{m}$.}
	\label{fig7}
\end{figure*}

\bibliography{Simionato_et_al_2020_supplementary}

\title{Methods}
\maketitle

\section{Sample preparation for training data generation}
Blood was drawn with informed consent from $5$ healthy donors via finger prick blood sampling in tubes containing $5\,\mu\text{l}$ of $1.6\,\text{mg/ml}$ EDTA. RBC shape transitions within the SDE scale were induced by osmolarity variation of the suspension solution (Supplementary Fig. 6). Ten microliters of each sample was resuspended in $1\,\text{ml}$ of NaCl solutions of different concentrations: $0.9\,\%$ NaCl was used to preserve discocyte shapes and $0.4\%$ NaCl was used to induce spherostomatocytes formation. Intermediate shapes were obtained by suspending cells in $0.5\%$ NaCl (stomatocyte types I and II) and $2.5\%$ NaCl (echinocyte types I and II) solutions (for echinocyte type III, see below).

A total of $400\,\mu\text{l}$ of each cell suspension was fixed in $1\,\text{ml}$ of $1\%$ glutaraldehyde (Sigma-Aldrich, Saint Louis, USA) solution in NaCl. To fix cells with the desired shape, each glutaraldehyde solution was prepared to have a total osmolarity equal to the osmolarity of the NaCl solution used to induce each shape \cite{abay2019glutaraldehyde}. Fixed cells were placed in a rotator mixer (Grant-bio PTR-35, Grant Instruments, Cambridge, England) at room temperature overnight. Samples were later centrifuged at $4000\,\text{g}$ for $5$ minutes (Eppendorf Micro Centrifuge 5415 C, Brinkmann Instruments, NY, USA), washed $3$ times with $1\,\text{ml}$ of each respective NaCl solution used to induce the different shapes and eventually resuspended in $1\,\text{ml}$ of the same solution. Five microliters of CellMask\texttrademark Deep Red plasma membrane stain $0.5\,\text{mg/ml}$ (Thermo Fisher Scientific, Oregon, USA) was added to each sample for $24$ hours at room temperature.

Then, the cells were washed $3$ times as described previously. Echinocyte type III was formed with live cells by exploiting the glass effect. First, $10\,\mu\text{l}$ of blood was suspended in $1\,\text{ml}$ of PBS and labeled with CellMask\texttrademark Deep Red. After washing, the cells were resuspended in PBS and finally placed on a glass slide for confocal microscopy. For later testing, $5\,\mu\text{l}$ of fresh blood drawn by finger pricks from $10$ volunteers and $10$ patients with hereditary spherocytosis was fixed in $1\,\text{ml}$ of either $1\%$ or $0.1\%$ glutaraldehyde in PBS, stained as described above and imaged. Samples from patients were fixed and shipped from Fondazione IRCCS Ca' Granda Ospedale Maggiore Policlinico of Milan (Italy) and University Medical Center Utrecht (The Netherlands).

\section{Imaging by confocal microscopy}
Each labeled sample was placed between two glass slides for imaging (VWR rectangular coverglass, $24\times 60\,\text{mm}$) on top of a $60\,\text{X}$ objective (CFI Plan Apochromat Lambda $60\,\text{X}$ Oil, $\text{NA} = 1.4$, Nikon, Tokyo, Japan) of an inverted microscope (Nikon Eclipse Ti). A solid-state laser ($\lambda = 647\,\text{nm}$, Nikon LU-NV Laser Unit) was used as a light source for imaging. Z-stack scanning was realized by employing a $300\,\text{nm}$ piezo stepper for a $20\,\mu\text{m}$ z-range. Confocal image generation was performed with a spinning-disk based confocal head (CSU-W1, Yokogawa Electric Corporation, Tokyo, Japan). Image sequences were acquired with a digital camera (Orca-Flash 4.0, Hamamatsu Photonics, Hamamatsu City, Japan).

\section{Image preprocessing}
A custom written MATLAB\texttrademark (MathWorks, MA, USA) routine was used to crop single cells from each image and perform their 3D reconstruction to enable visualization of the 3D shape. Each single-cell 3D image contained $68$ individual planes with an extent of $100\,\text{px}\,\times\,100\,\text{px}$ and a lateral (x/y) resolution of $0.11\,\mu\text{m/px}$. The piezo stepper had a minimal step width of $0.3\,\mu\text{m}$, defining the z-resolution accordingly. To compensate for the difference in resolution in the x/y  and z directions, we adapted the scale in z by means of linear interpolation. Thus, the obtained z-stack had dimensions of $100\,\text{px} \times 100\,\text{px} \times 185\,\text{px}$. The image stacks were then passed to a custom written ImageJ script. By applying a fixed threshold for every image, the script binarized the confocal z-stack to retrieve the cell membrane as an isosurface. After vectorization, the origin of the cell always corresponded to the center of its bounding box. Therefore, this step had the benefit of introducing an inherent translation invariance, i.e., for a given rotation, the form and size of the bounding box is invariant regarding translations in space. The obj-files (Wavefront Technologies, CA, USA) generated in such a manner were then automatically transformed into the polygon file format (ply) and passed to the shape descriptor for the spherical harmonics analysis.

\section{Spherical harmonics analysis}
The spherical harmonics analysis was performed by using the high-performance software implementation described by Kazhdan et al. \cite{kazhdan2003rotation}. The algorithm first maps a given 3D object (ply format) onto a 3D voxel grid of defined size. For our purpose, we kept the standard parameters of the algorithm, using a voxel grid of $64\,\text{px}\times 64\,\text{px}\times 64\,\text{px}$. Within this cube, at least $32$ spherical functions of different radii can be arranged around the center point. Each spherical function can then be decomposed as the sum of its harmonics:

\begin{equation}
f\left(\Theta,\phi\right) = \sum_{l=0}^{\infty}\sum_{m=-l}^{m=l}a_{lm}Y_{l}^{m}\left(\Theta,\phi\right)
\end{equation}

Keeping the standard settings, for every radius, $16$ frequencies (harmonics) were calculated. In addition, the first- and second-order components of each decomposition were expressed in a Euclidean manner using $3$ scalars $a_{1},a_{2}$, $a_{3}$:

\begin{equation}
f_{0} + f_{2} = a_{1}x^{2} + a_{2}y^{2} + a_{3}z^{2}.
\end{equation}

In the case of principal component analysis (PCA), these values could potentially be used for alignment purposes. However, in our case, these factors were not employed for further PCA investigations but were included in the training process. After transformation, each frequency component was accumulated by calculating the corresponding $L_{2}$-norm. This resulted in a vector of size $32 \times (3 + 14) = 544$, where $32$ corresponded to the number of radii, $3$ to the scalars describing $f_{0}$ and $f_{2}$, and $14$ to the number of remaining frequencies for each radius. Finally, each cell was expressed by means of a one-dimensional vector. For further signal processing, this vector was normalized and mapped onto a numeric range from $0$ to $1$. The intrinsic rotation invariance with respect to the original 3D data was the key factor in choosing this kind of transformation for shape description. This is because cells can have any orientation after sedimentation on the microscopy slide, and a general expression of their shape is required for further analysis.

\section{Manual classification}
Each vectorized cell was rendered using Blender for cell shape visualization to perform a manual shape classification. The samples fixed in solutions at different osmolarities were used to classify the various shapes of the SDE scale based on Bessis' criteria \cite{lim16766, larkin2013nmr}: (1) discocytes, meaning biconcave and symmetric disks; (2) stomatocytes type I and II, characterized by a lighter (I) and deeper (II) monoconcavity; (3) spherocytes, i.e., spherical cells; (4) echinocyte type I, crenated cells preserving biconcavity; (5) echinocyte type II, cells with forming spicules; and (6) echinocyte type III, cells with more than 25 spikes. Following the observation of other cell morphologies, the chosen shapes for additional categories were knizocytes, i.e. trilobal cells resulting from high shear stress or observed in some blood diseases; keratocytes, a category including damaged RBCs with variable shapes; acanthocytes, including echinocytes with irregular spicules and/or a spherical body; and multilobate cells, i.e., young reticulocytes. A class for the exclusion of an artifact occurring upon cell fixation, named cell clusters, was added. Each class contained a minimum of $10$ cells to a maximum of $200$ cells. Any other shape beyond the chosen classes was not considered and therefore not introduced in the training process.\\

\section{Training and validation of the dual-stage ANN}
Keras with TensorFlow as the backend was used to build and train the dual-stage ANN. A set of representative RBC shapes was chosen for the supervised training of both ANN stages. Each of the selected cells was then transformed according to the previously discussed steps. In addition, data augmentation of the spectra was performed by creating $1000$ linear interpolations between randomly picked spectra belonging to the same class, in the case of the first-stage ANN (classification) and between spectra of neighboring pseudodiscrete classes in the case of the second-stage ANN (regression) to cover the whole SDE shape spectrum. The augmentation also served to compensate for the different number of ideal cell shapes that were found for certain classes, ensuring a balanced training dataset (1000 total data per class). Similar to the k-fold cross validation approach, we trained both ANN stages with $100$ different random starting conditions, finally selecting the best-performing ANN for each stage. The training data constituted $80\,\%$ of the whole set of data, while the remaining $20\,\%$ was used for ANN validation. 
The training was performed in batches of $100$ spectra for both ANNs, finalizing the process at $100$ and $40$ epochs for the first-stage and second-stage ANN, respectively. 
The related loss functions were the crossentropy for the first and the mean squared error (MSE) for the second ANN, while the chosen optimizer for both was Adam \cite{kingma2014adam}.

\section{Automatic classification of 3D shapes in healthy individuals and patients}
Samples from $10$ healthy subjects and $10$ patients affected by hereditary spherocytosis were automatically classified by the dual-stage ANN. The number of tested cells per sample ranged from about $1000$ to over $2000$ cells. The results were manually verified in Blender. Samples from $5$ healthy donors and $5$ patients were chosen for the manual selection of $200$ cells each for comparison with the automatic shape recognition and were plotted in a related confusion matrix.

\section{Blood smear}
Five milliliters of whole blood was drawn in EDTA anticoagulant. Thin peripheral blood smears were set up within $4$ hours from blood sampling, dried and stained according to the May Gr\"unwald-Giemsa method \cite{bain2012practical}. RBC morphology and the number of spherocytes were independently assessed by two laboratory experts. The number of spherocytes was expressed as a percentage of the total number of RBCs observed.\\


%\bibliography{Simionato_et_al_2020_methods}